**Recent advances in terahertz photonic technologies based on graphene and their applications**

*Tianjing Guo and Christos Argyropoulos\**

Prof. Tianjing Guo

Institute of Space Science and Technology, Nanchang University, Nanchang 330031, China

Prof. Christos Argyropoulos (*corresponding author*)

Department of Electrical and Computer Engineering, University of Nebraska-Lincoln, Lincoln, Nebraska 68588, USA

E-mail: christos.argyropoulos@unl.edu

**Abstract**

Graphene is a unique two-dimensional (2D) material that has been extensively investigated owing to its extraordinary photonic, electronic, thermal, and mechanical properties. Excited plasmons along its surface and other unique features are expected to play an important role in many emerging photonic technologies with drastically improved and tunable functionalities. This review is focused on presenting several recently introduced photonic phenomena based on graphene, beyond its usual linear response, such as nonlinear, active, topological, and nonreciprocal effects. The physical mechanisms and various envisioned photonic applications corresponding to these novel intriguing functionalities are also reported. The presented graphene-based technologies promise to revolutionize the field of photonics at the relatively unexplored terahertz (THz) frequency range. They are envisioned to lead to the design of compact harmonic generators, low-power wave mixers, linear and nonlinear sensors, magnet-free isolators and circulators, photonic topological insulators, modulators, compact coherent optical radiation sources, and subwavelength imaging devices.

**1. Introduction**

Graphene is composed of a monolayer of carbon atoms arranged in a 2D honeycomb lattice. It was first isolated in its planar form in 2004 by using an adhesive tape.[1] Since its discovery, graphene has become one of the most investigated 2D materials because it combines remarkable photonic, electronic, thermal, and mechanical properties.[2–4] The high electrical and thermal conductivity of graphene can be attributed to its unusually large room temperature electron mobility of approximately 15,000 cm$^2$/Vs.[5,6] In addition, surface plasmons can be formed along its surface at THz frequencies when excited by external optical radiation, an advantageous property that can lead to a plethora of new compact THz photonic devices.[7] The adhesive tape exfoliation is not practical for large-scale photonic device fabrication but, more recently, wide-area and uniform 2D graphene monolayers were synthesized by using the chemical vapor deposition (CVD) method.[8,9]

Graphene can be patterned into different shapes, such as micro/nanoribbons and disks,[10–12] which can be used as the building elements of new ultrathin graphene metasurface designs.[13] Graphene metasurfaces exhibit localized plasmon resonances and, as a result, stronger light-graphene interactions at THz frequencies compared to unpatterned graphene monolayers. These types of metasurfaces have attracted increased attention because of their various THz applications, such as broadband absorbers, modulators and filters, biosensors, tunable polarizers, light detectors, and cloaking devices.[10,14–29] Furthermore, graphene is perfectly suited to be combined with metallic (plasmonic) or dielectric metamaterial or metasurface structures, cavities, and gratings due to its atomic-scale thickness and conformal nature.[14,16,29–

[33] The hybrid graphene-metamaterial design approach promises to further enhance the efficiency of light-graphene interactions. More importantly, it can lead to the design of tunable compact photonic devices by changing graphene's electromagnetic properties via chemical doping or electrical gating.[32,34–36] As an example, enhanced and tunable THz transmission modulation was reported by coupling graphene with dielectric or plasmonic metasurfaces.[29,37]

Graphene has also been predicted to possess very strong optical nonlinear response, especially at THz frequencies, due to its peculiar Dirac cone band structure, as shown in **Figure 1**.[38] Its nonlinear effects can be triggered by rather low electric field values (~$10^3$ V/cm) of the external incident wave.[39–41] The strong nonlinear response of graphene arises from the fact that its carrier velocity is not proportional to momentum under an oscillating electromagnetic field. To make this more clear, while in usual parabolic band semiconductors, velocity ($v$) and momentum ($p$) are proportional and related by the expression: $v_x = \partial \varepsilon(p)/\partial p_x \propto p_x$, in graphene this relation is different: $v_x = \partial \varepsilon(p)/\partial p_x \propto p_x/\sqrt{p_x^2 + p_y^2} \propto sgn[p_x]$,[42] where $\varepsilon(p)$ is the energy spectrum of the quasi-free electrons in graphene.[39] The harmonic decomposition of the "sgn" function can lead to the generation of all odd high-harmonics, because $p_x$ is a sinusoidal function, causing efficient frequency multiplication and high harmonic generation.[39] Hence, graphene has been experimentally demonstrated to possess a remarkably strong third-order nonlinear susceptibility $\chi^{(3)}$ at THz and IR frequencies due to its unique Dirac cone band structure shown in Figure 1.[42–46]

Graphene can also provide a promising platform for the realization of other novel photonic technologies in THz frequencies, such as compact topological and nonreciprocal photonic systems.[47–50] In particular, it becomes gyrotropic material under magnetic bias[51] or nonreciprocal medium in the case of spatiotemporal modulation.[49] The envisioned nonlinear, nonreciprocal, active, and topological graphene-based photonic devices consist viable new ways to introduce groundbreaking technological advances in THz frequency range.[52] While linear and passive graphene-based devices have been extensively reviewed before,[34,53–57] the emerging nonlinear,[58,59] active, topological, and nonreciprocal properties of graphene and their resulted cutting-edge technological systems have not been discussed in detail during previous related review works.

All these intriguing new graphene functionalities are reviewed and presented in our paper that is organized in five sections. Following the introduction in section 1, section 2 provides a detailed description of graphene's strong nonlinear response. It is demonstrated that patterned graphene, forming graphene metasurfaces, and the hybridization of graphene with plasmonic or dielectric structures constitute the two main approaches to design more practical THz nonlinear devices and further boost graphene's large intrinsic nonlinearity. The active response of photoexcited graphene is presented in section 3. Ways to achieve efficient negative conductivity or gain when graphene is optically pumped are also discussed in the same section, as well as the intriguing THz devices that can be realized by this active response. Section 4 deals with the recently emerged topological and nonreciprocal photonic devices based on graphene and their physical mechanisms and implementations. Finally, several conclusions and future perspectives on THz photonic technologies based on graphene and beyond are provided in section 5.

## 2. Nonlinear graphene

Doped graphene has been found to possess strong nonlinear electromagnetic properties[40,60] that can be described by its high nonlinear susceptibility. This parameter is used to quantify the nonlinear property of a material by determining its nonlinear polarization in terms of the strength of an applied electric field.[61] The general linear and nonlinear polarization formula is $\tilde{P}(t) = \varepsilon_0[\chi^{(1)}\tilde{E}(t) + \chi^{(2)}\tilde{E}^2(t) + \chi^{(3)}\tilde{E}^3(t) + \cdots]$, where $\varepsilon_0$ is the permittivity of free space, $\chi^{(1)}$ is the linear susceptibility, and $\chi^{(2)}$ and $\chi^{(3)}$ are the second- and third-order nonlinear susceptibilities, respectively. Note that the second-order nonlinear response of a free-standing graphene monolayer is extremely weak within the electric dipole approximation because graphene has centrosymmetric properties.[62] However, the inherently weak second-harmonic generation (SHG) from graphene can be enhanced by coupling to plasmonic structures[63,64] or taking into account quadrupolar resonances or spatial dispersion.[65–67] Graphene was experimentally demonstrated to possess a remarkably strong $\chi^{(3)}$ at THz frequencies, which originates from the intraband electron transitions (see section 1 for more details), as well as the excitation of electrically tunable plasmons[7,34] that lead to enhanced electromagnetic fields with extreme surface confinement.[42,68] Specifically, the Kerr nonlinear susceptibility of graphene was found to reach high values ($1.4 \times 10^{-15} m^2 V^{-2}$) in experiments performed at IR frequencies.[46] In addition, the nonlinear response of graphene was calculated in the THz frequency range by developing a theoretical model based on the density-matrix formalism, demonstrating that the third-order nonlinearity of graphene can be further improved by optimizing its Fermi level.[69] Even larger record-breaking effective third-order nonlinear coefficients of $\chi^{(3)} \approx 10^{-9} m^2 V^{-2}$ were measured in recent experiments at low THz frequencies.[70] These large nonlinear coefficient values make graphene an ideal material platform to be used in frequency generators or other compact THz optoelectronic applications.

On a related note, plasmonic configurations have been widely demonstrated to exhibit very high enhancement of various nonlinear effects at the IR and visible spectrum, due to their enhanced and confined electromagnetic fields when surface or localized plasmons are excited.[71–83] However, the nonlinearity enhancement in THz frequencies based on plasmonic or dielectric structures still remains rather limited mainly due to the high reflective nature of metals and the low field confinement of dielectrics in this frequency range. Graphene promises to solve this problem. More specifically, the hybrid approach of combining graphene monolayers with plasmonic or dielectric gratings, cavities, and metamaterials seems to be promising to further boost nonlinear effects at THz frequencies and make them more practical.[84] As an example, the SHG response of graphene was enhanced through its resonant coupling to a plasmonic metasurface composed of a lattice of asymmetric gold split-ring resonators (SRRs) exhibiting a Fano resonance.[63]

Note that THz surface plasmons in graphene monolayers face the usual energy and momentum mismatch problem with the incident light in free space, similar to optical surface plasmons on metallic (plasmonic) interfaces.[85] Hence, special coupling techniques, such as near-field excitation, need to be applied to excite THz surface plasmons on graphene monolayers. This problem ceases to exist by using patterned graphene configurations, since the geometrical features will change the dispersion equation of graphene plasmons and make them accessible to the incident light without the need of additional coupling mechanisms. Hence, it is favorable to fabricate new graphene structures by patterning a continous graphene sheet to periodic ribbons, disks, rings, or triangles, to more efficiently excite plasmons that will further enhance the nonlinear response.[86–89] For example, by arranging graphene in ribbons or patches to form metasurfaces, a large enhancement in the effective second-order susceptibility by more than three orders of magnitude was realized compared to the intrinsic second-order susceptibility of an unpatterned graphene monolayer placed on the same glass substrate.[90] This advantageous effect was mainly due to the double-resonant nature of these graphene metasurfaces at the fundamental and second harmonic frequency. At this point, it is relevant to note that the centrosymmetry SHG contraint is broken when graphene is placed on a substrate. By terminating the graphene ribbons on glass substrate with a gold reflector (design shown in **Figure 2**(a)), an even larger boost in the third-order nonlinear process of third-harmonic generation (THG) was achieved. In this particular case, the THG output power was dramatically enhanced by several orders of magnitude compared to the case of no reflector, unpatterned graphene or no graphene,[91] as illustrated in Figure 2(b). The comparison results revealed that the graphene ribbons combined with the gold reflector played an important role in triggering the efficient THG process. In addition, graphene metasurfaces consisting of a rectangular array of cruciform graphene patches also exhibited enhanced effective third-order nonlinear susceptibility.[92] The nonlinear homogenization method was employed in some of the aforementioned works,[90,92] making it possible to replace the patterned graphene metasurface by a simpler homogenous material layer characterized by an effective permittivity and nonlinear susceptibility retrieved by the homogenization method. Localized plasmons due to graphene nanoribbons were also reported to control the nonlocal nature of the resulted effective third-order nonlinearity leading to efficient light mixing generation.[93] Moreover, in a relevant design, the patterning of graphene to ribbons led to strong THz nonlinear absorption, enhanced by two orders of magnitude compared to unpatterned graphene, directly indicating the plasmon-enhanced nonlinearity.[94]

Other forms of graphene scatterers, such as graphene nanoislands, were also investigated to achieve localized plasmons. Their derived nonlinear polarizabilities were electrically tuned to surpass those of metal nanoparticles of similar size by several orders of magnitude.[89] Nonlinear polarization at multiple harmonics was produced under the illumination of a polarized light pulse on these graphene nanoscatterers leading to efficient SHG and THG processes. Moreover, the nonlinear optical wave mixing process based on the nanoisland design was also investigated,[95] as schematically illustrated in Figure 2(c). Extraordinary high wave-mixing susceptibilities were obtained by the doped graphene nanoislands when one or more of the input or output frequencies coincided with their multiple pronounced localized plasmon resonances. The wave mixing enhancement was tunable over a wide frequency range spanning visible and IR, just by controlling the doping level of the nanoisland without changing its geometry.

As was mentioned before, an alternative approach to boost the intrinsic high nonlinearity of graphene is by hybridizing it with metamaterials, cavities or plasmonic gratings. Towards this goal, it was reported that the hybridization of graphene with asymmetric plasmonic SRRs can enhance by more than an order of magnitude the transient nonlinear transmission response of bare graphene within a broadband frequency range.[96] Another recent experimental work based on a hybrid metamaterial made of a metallic grating combined with graphene is depicted in **Figure 3(a)**.[97] In this experiment, a narrow-band THz waveform was focused on the sample and the electric field of the transmitted waveform was measured as a function of time. The harmonic generation of the proposed metamaterial structure was demonstrated by comparing the field strength of the transmitted signal to that of incident light in frequency domain by computing the Fourier transform of the resulted time domain signals. Figure 3(b) illustrates the THG nonlinear properties of the two samples quantitatively. It was found that the THG of the hybrid metamaterial is more than an order of magnitude higher compared to the bare graphene under low incident field strength, while THG is similar for the two samples when the incident field strength is high. The quantity shown by the x-axis in Figure 3(c), named duty cycle, is defined as the ratio of metal width to the period of the metallic grating. It was shown that the nonlinear conversion efficiency can be improved by increasing the duty cycle. This was because a larger duty cycle means a smaller gap in between the metal stripes that leads to larger field enhancement. In addition, intense higher order harmonics, beyond the third harmonic, were also generated by the hybrid metamaterial with results shown in Figure 3(d). These findings illustrate that the ninth-harmonic intensity of the proposed grating-graphene hybrid metamaterial can be enhanced by more than seven orders of magnitude compared to bare graphene. Hence, high THz nonlinear harmonic conversion efficiency was experimentally achieved by low power consumption.

Furthermore, a relevant study based on a metallic grating-graphene hybrid metamaterial is demonstrated in **Figure 4(a)**. In this case, the third harmonic was generated in reflection, which led to even higher nonlinear conversion efficiencies.[98] This structure was composed of a conventional plasmonic grating with graphene placed on top of it. High THG conversion efficiency was obtained (~18%) by using low input intensities, benifiting from the robust localization and enhancement of the electric field within the trenches of the plasmonic grating. Different third-order nonlinear processes, as shown in Figures 4(a) and 4(c), were found to exhibit an impressive enhancement of more than twenty orders of magnitude compared to the same plasmonic grating structure but without graphene. By calculating the third harmonic and four-wave mixing (FWM) output power under different scenarios, as shown in Figures 4(b) and 4(d), respectively, it was demonstrated that the THG and FWM output powers of the proposed graphene-covered hybrid metamaterial grating are much higher compared to the cases when graphene was not present.[98] It was also found that the nonlinear response remains relative insensitive across a broad range of incident angles in this configuration. The enhanced omnidirectional nonlinear response of these hybrid graphene-plasmonic structures can be used in several promising applications, including THz frequency generators, all-optical signal processors, wave mixers, as well as nonlinear THz spectroscopy and noninvasive THz subwavelength imaging devices.

Except of enhanced THG, several other nonlinear effects based on graphene's third-order nonlinearity have been explored in the literature, such as optical bistability, Kerr nonlinear process, nonlinear refraction and saturable absorption, and wave-mixing.[36,93,94,99,100] More specifically, optical bistability was demonstrated by a graphene monolayer or patterned graphene ribbons.[99,100] This nonlinear response can be useful in THz optical switching applications. In addition, strong nonlinear frequency mixing of incident light was generated by an extended graphene monolayer with the experimental arrangement shown in **Figure 5**(a).[101] Surface plasmons with a defined wavevector and direction were excited all-optically along graphene in this configuration by taking advantage of graphene's high intrinsic nonlinear optical property. The lower branch of the graphene plasmon dispersion relation led to the largest mixing signals due to the low-wavevector phase matching, while the upper dispersion branches caused the strongest signals to be generated at the long-wavevector range with results demonstrated in Figure 5(b). Figure 5(c) presents the experimental set-up and schematic of a nanostructured graphene geometry that generated efficient light mixing due to the excitation of localized plasmons.[93] The generated signal at the combination frequency is plotted in Figures 5(d) and 5(e) in time and frequency domain, respectively. In this case, it was reported that the measured plasmon-assisted nonlinear conversion efficiency from the graphene nanostripes was two orders of magnitude larger than that measured from a continuous graphene monolayer,[46] where no excited localized plasmons exist due to the lack of nanopatterning. The envisioned potential applications of this efficient wave mixing scheme include THz multiplexers, modulators, and sensors.

Extremely high THz high-harmonic generation (up to the seventh-order) was achieved by utilizing the thermodynamic response of a graphene monolayer free electrons.[70] In this interesting recent experiment, a CVD-grown graphene sample was deposited on a $SiO_2$ substrate that was excited by a quasi-monochromatic high power pulse with frequency 0.3 THz, as depicted in the schematic of Figure 5(f). Enhanced odd-order harmonic generation was measured in the transmitted THz spectrum shown in Figure 5(g). Remarkably strong harmonic conversion efficiencies were achieved by this experiment with extremely high values around $10^{-2}$, $10^{-3}$, and $10^{-4}$, for the third, fifth, and seventh THz harmonic wave, respectively. It was concluded that the nonlinear optical response of graphene at THz frequencies is orders of magnitude stronger compared to conventional bulky nonlinear materials. In this work, the thermodynamic analysis of graphene under high power THz illumination was used to predict its nonlinear THz conductivity.[102]

## 3. Active graphene

Graphene's optical loss can hinder many of its practical applications. In particular, the large positive real part of graphene surface conductivity, that characterizes its losses, makes the propagating THz plasmons to quickly decay and the local resonant plasmonic modes to be weakly excited in graphene-based plasmonic devices. [103][38,104,105] Loss compensation by using a gain medium can be considered a straightforward strategy to further improve and enhance light-matter interactions along the presented tunable graphene photonic structures. However, graphene can offer a novel mechanism of inherent loss compensation by exhibiting population inversion due to its nonequilibrium carrier dynamics[106–108] when it is optically pumped by external radiation with arbitrary energy $\hbar\Omega_0$, as shown in **Figure 6.**[108] The population inversion is demonstrated by calculating the dynamic conductivity of pumped graphene, where its real part becomes negative in the THz frequency range, as shown in Figure 6(b).[106,109–111] Figures 6(c)-(e) illustrate the carrier relaxation and recombination temporal dynamics in optically pumped graphene from few tens of femtoseconds (fs) to picoseconds (ps) after the pumping process. Within the 20–200 fs time scale regime (Figures 6(c)-(d)), collective excitations play a dominant role causing an ultrafast carrier quasi-equilibrium to be reached due to the intraband femtosecond-scale carrier–carrier scattering. Then, the carriers at their high-energy distribution tails emit optical phonons due to the fast interband relaxation process, hence, cooling themselves and accumulating around the Dirac points, which is demonstrated in Figure 6(e). Due to the fast intraband relaxation time and relatively slow interband electron-hole recombination process, an efficient population inversion is achieved (Figure 6(e)) that can lead to THz lasing.[112] Note that the thermodynamic analysis of graphene under optical pump excitation can also be used to understand the physics of its negative real part transient conductivity or gain.[113] This intriguing process can also be realized by carrier injection besides the aforementioned optical pumping approach.[114,115] Active optically pumped graphene can be integrated within resonators, nanocavities, metamaterials, or waveguides to design high-efficient transient THz laser sources[116,117] that will be used for sensing, modulation, and imaging applications.

Understanding the dynamics of graphene's optical response to photoexcitation is an interesting fundamental science problem but is also critically important to the development of low-loss high-performance THz optoelectronic passive and active devices. It was reported that the real part of the THz conductivity can become negative at optically pumped graphene, providing intriguing possibilities to design THz laser sources generating long-wavelength coherent radiation.[106] This was a significant research milestone with the potential to dramatically alter THz technology that currently lacks high power coherent radiation sources mainly due to the absence of suitable gain media in this frequency range.

This interesting effect was further experimentally explored by measuring the interband electron-hole recombination rates of graphene epitaxially grown on a silicon carbide (SiC) substrate by using optical pump THz-probe spectroscopy.[107] It was found that the transient conductivity of graphene highly depends on the carrier concentration and energy distributions. The unusual transient properties of photoexcited graphene were further confirmed by measuring the ultrafast minority-carrier nonequilibrium recombination time dynamics via detecting changes in its THz transmission.[118] These early works demonstrated increased transient conductivity values for graphene upon ultrafast optical illumination. A pumped-induced decrease in the graphene conductivity was reported by observing the transient THz transmission response from single-layer CVD-grown graphene samples with relative high mobility and doping level values.[119] This motivating result was in agreement with another relevant study,[120] where it was demonstrated that in highly doped graphene the photoexcitation has no effect on carrier concentration but increases the electron scattering rate that, subsequently, decreases its conductivity.[121]

In another relevant experiment, the evolution of hot carrier relaxation dynamics at different graphene Fermi energies or doping levels were revealed.[122] The predominant scattering source in this study was credited to the charged impurities of the graphene that was grown by CVD on a quartz substrate.[122] Recently, the correlation between the transition from positive to negative THz photoconductivity was attributed to two types of ultrafast photo-induced carrier processes: interband and intraband heating.[123] The positive (negative) photoconductivity at low (high) Fermi energies was caused by the effect of the heated carrier distributions on the screening of impurities. Hence, positive photoconductivity was found to result from a rise in the carrier concentration, caused by an increase in the Fermi level, when the graphene sample was close to its neutral charge point. On the contrary, negative photoconductivity occurred only in highly doped samples and was connected to the relaxation of the photoexcited carriers through many mechanisms, such as the increase of carrier temperature, enhanced carrier-carrier scattering, and boosted carrier scattering with lattice vibrations. A model for the THz and IR photoconductivity of graphene at room temperature was lately developed.[124] This model is the most accurate up to now, since it includes both energy relaxation and generation-recombination associated with optical phonons and Auger recombinations that are crucial processes to accurately determine the photoconductivity response.

The negative conductivity of optically pumped graphene can directly lead to the realization of the elusive THz lasing source. Several designs of THz coherent sources based on population inversion from graphene were proposed in the past.[110,114,116,117,125–128] The dominant process of strong nonradiative carrier recombination was found in the majority of the studies to be the key obstacle to the appearance of pronounced optical gain in graphene. To address the intriguing question whether a long-lived strong optical gain can be achieved in graphene, it was proposed to integrate a graphene monolayer into a high quality factor photonic crystal cavity to further boost its carrier-light interactions.[126] Additionally, a high-dielectric substrate was used below graphene to reduce the efficiency of Coulomb-induced nonradiative recombination processes. The schematic of this proposal, along with the dominant mechanisms in optically pumped graphene, are shown in the top and bottom captions of **Figure 7**(a), respectively. After optical excitation, the nonequilibrium carriers relax to lower energies via carrier-carrier and carrier-phonon scattering.

Another recent interesting design to realize THz lasing was proposed to be made of patterned hyperbolic metamaterials[129–132] composed of multiple stacked photoexcited graphene layers separated by thin dielectric sheets, as depicted in the inset of Figure 7(b).[133] The patterning of the hyperbolic metamaterial structure was found to support slow-wave modes in the THz region that made it possible to drastically boost the THz wave amplification. Strong gain was achieved by this configuration, as depicted in Figure 7(b), which is expected to lead to an efficient compact THz laser design. This active graphene-based hyperbolic metamaterial device was also demonstrated to be ultrasensitive to its surrounding environment, making it ideal candidate for biological sensing applications with the ultimate goal to improve the diagnostics of several diseases.

In general, 2D materials exhibit extreme sensitivity to their dielectric environment. The same is true for graphene, because of its ultrasensitive static and dynamic electrical characteristics.[86,134] Recently, an alternative intriguing approach to sensing by using an optically-pumped active graphene device based on the parity-time (PT) symmetry concept was proposed by using the design shown in **Figure 8**(a).[135] The property of the negative real part conductivity in photoexcited graphene was crucial to achieve this PT-symmetric system that exhibited gain, caused by the pumped graphene metasurface, balanced by the loss from the metallic filament. Very large sensitivity to an extremely low-level change in graphene's chemical doping triggered by the surrounding environment was predicted by the computed transmission and reflection results plotted in Figure 8(b).[135] Note that a similar passive graphene plasmonic sensor design demonstrated substantially reduced sensitivity compared to the performance of the presented active graphene PT-symmetric THz sensor. This active graphene sensor can have a substantial tunable transmission/reflection performance by varying in a larger scale its chemical doping, for example by electrostatic gating.

Active photoexcited graphene can also be used to design efficient THz modulators due to loss compensation.[136] For instance, the photoexcited graphene metasurface presented in **Figure 9**(a) was used to achieve an efficient low-loss transmission modulator.[137] The monolayer graphene was patterned into a periodic array of SRRs to form a metasurface that exhibited a magnetic resonant response. The resonantly enhanced energy dissipation of this magnetic response was compensated by the photoexcitation of graphene leading to a sharper resonance behavior as the quasi-Fermi level was increased from 0 to 10.3 meV, as illustrated in Figure 9(b). The highest Q-factor transmission resonance was obtained at 1.5 THz when the quasi-Fermi level was 10.3 meV, which means that loss was almost completely compensated from the optically induced graphene gain at this point. Furthermore, it was found that the transmission modulation depth is greatly improved by increasing the quasi-Fermi level in this photoexcited magnetic metasurface and can reach very high values of approximately 90%. This work is expected to pave the way to the design of low-loss active magnetic resonators and efficient ultrathin planar transmission modulators at THz frequencies.

The relevant experimental study of a graphene-based thin film absorption modulator was recently reported.[138] It was demonstrated that it can indeed exhibit ultrafast absorption modulation at THz frequencies by applying an optically pumped signal on graphene.[138] **Figure 10**(a) illustrates the active absorber design that consists of a uniform CVD-grown graphene monolayer on a ground dielectric substrate terminated by a metallic ground plate. Terahertz time-domain spectroscopy was used in this experiment to measure the tunable absorption of this device. The system was also illuminated by a variable intensity IR pump laser to achieve the photoexcitation of graphene. It was derived that the graphene-based structure absorbs a maximum of 75% of the incoming THz radiation at the resonance frequency of 2.17 THz when graphene was not optically excited (fluence 0 mJ/cm$^2$), as demonstrated in Figure 10(b). The graphene's absorption was reduced to 45% when photoexcitation was applied by increasing the fluence to 0.690 mJ/cm$^2$. Ultrafast modulation of the absorption on the order of 40% was experimentally achieved with this photoexcited graphene absorber configuration. Moreover, there was a secondary absorption resonance peak around 6.38 THz, as can be seen in Figure 10(b), where the absorption modulation was lower and equal to 25% after photoexcitation of the graphene monolayer. The absorption amplitudes at these two resonant points as a function of the IR laser fluence are plotted in Figure 10(c). The absorption peaks amplitudes were decreased as the fluence was increased, which was attributed to the decrease in the real part of the graphene conductivity (loss compensation) due to the photoexcitation. This compact active graphene device can provide ultrafast and strong absorption modulation, ideal for several future flat optics THz modulation applications.

**4. Topological and nonreciprocal graphene**

Topology is a branch of mathematics that deals with highly conserved 'topological invariant' quantities that do not change when physical objects are continuously deformed. The implementation of topology in photonics promises to translate the physics of topological phases of matter in a novel optical context.[139,140] The ultimate goal of this emerging research field will be to achieve new strongly correlated states of photons with topological features that can lead to a novel generation of photonic devices, such as unperturbed waveguides, topological lasers, and optical isolators. Recently, the emerging field of topological photonics has attracted increased research interest mainly because of the new route it provides to make the propagation of light robust against disorder and backscattering.[139–144] A wide range of geometries, such as photonic crystals, metamaterials, cavities, and waveguides,[145–154] have been reported to demonstrate various topological photonic effects. The dynamic modification of topological photonic phenomena is expected to increase the practical applications of this innovative concept. For instance, the dynamic control of topological phase transitions was achieved by immersing a topological photonic crystal into a liquid crystal background.[155] Reconfigurable topological photonic systems were also reported when using a bianisotropic photonic crystal.[156] These interesting recent findings can provide a robust approach towards the dynamic tuning of propagating waves along any desired path without the adverse effects of back-reflection and scattering. On a related note, self-induced optically tunable topological protected electromagnetic radiation was demonstrated with a nonlinear photonic crystal composed of a patterned silicon slab under optical pumping.[157] The concept of topological photonics was also investigated to achieve stable and low-threshold lasing by using the topologically protected edge states.[158–160] As an example, a recent work has experimentally demonstrated a high-performance topological laser based on a photonic crystal cavity.[161] In addition, it was found that large directional third-harmonic signal can be generated from a topologically non-trivial zigzag array of dielectric particles,[162] paving the way to novel topological radiation generation induced by nonlinearities.[163]

While all the aforementioned designs operate at microwave or near-IR/visible frequencies, graphene promises to extend these exciting topological photonic concepts to THz frequencies. Graphene is an ideal platform to study THz topological photonics due to the excitation of topologically protected plasmons when biased with magnetic field to break time-reversal symmetry.[142,164] Topological plasmonic states without the need of an external magnetic bias (preserved time-reversal symmetry) can also be achieved with graphene systems by varying graphene's doping level in particular areas.[165,166] These topological plasmons exhibit long propagation lengths and have a broadband nature mainly because of the large topological bandgap, which is enabled due to graphene's exceptional properties, such as long intrinsic relaxation time, large carrier densities, and ultrasmall Drude mass.[167–169] Moreover, miniaturization and tunability are two unique features of graphene topological photonic structures compared to other relevant designs based on bulk materials.

Hence, topological one-way edge states were realized by applying a static magnetic field bias to a periodically patterned graphene monolayer with geometry shown in **Figure 11**(a).[164] This design breaks time-reversal symmetry and leads to topologically protected edge plasmon modes propagating unperturbed along structural defects, as depicted in Figure 11(b). The operation frequency of these topological plasmons was tuned from THz to far-IR by suitably engineering the plasmonic band structure induced by the patterned geometry. Interestingly, large topologically induced bandgaps were maintained even under modest magnetic field bias values that can be tunable for varying graphene doping levels.[164] The alternative configuration of a 2D graphene photonic crystal shown in Figure 11(c) was investigated to achieve topological effects without magnetic bias. It consisted of an array of graphene nanodisks with different doping levels (chemical potentials) arranged in a honeycomb lattice on a dissimilarly doped graphene monolayer. Topologically protected edge states were also realized with this configuration by tuning the chemical potential of adjacent nanodisks, as depicted in Figure 11(c).[165] Dynamically gate-tunable topological plasmon modes (Figure 11(c)) were obtained by this design operating in a broad spectral range. The presented topological graphene-based devices are expected to bring the exciting new field of topological photonics to THz frequencies, hence creating efficient integrated THz nanowaveguides with an extremely compact profile and tunable response that can be used in the next generation communication devices.

The graphene-based designs presented in Figure 11 are linear topological systems. However, nonlinear topological photonic structures have started to attract increased attention mainly due to the self-induced tunability of their topological properties by changing the incident light intensity.[162,163,170] Graphene can play a pivotal role in this emerging field due to its large nonlinearity that can be tuned over a broad spectral range, as was described in the previous section 2. As an example, the topologically protected nonlinear FWM process was recently realized with a graphene metasurface upon breaking its time-reversal symmetry by a static magnetic field bias.[171] This nonlinear graphene metasurface design was made of periodic nanoholes with hexagonal symmetry located on a graphene monolayer with geometry depicted in **Figure 12**(a). Topologically protected edge modes were generated at both idler and signal frequencies due to the nonlinear FWM interaction. The relevant simulation results are shown in Figures 12(b) and 12(c), where edge modes are obtained at signal and idler frequencies, respectively. Both signal and idler modes are topologically protected and exhibit unidirectional unpertubed propagation along the edge of the same nonlinear graphene metasurface system, clearly demonstrating its ultrabroadband topological response.

The breaking of Lorentz reciprocity law is another relevant emerging photonic research area that can be impacted by graphene.[172] New compact nonreciprocal systems, down to atomic scale, operating at the THz frequency range can be realized by graphene due to its extraordinary properties. Specifically, graphene will become nonreciprocal and strongly gyrotropic under an external magnetic field bias.[173–175] This property was also utilized in the topological results presented before in Figures 10(a) and 11. The giant nonreciprocal properties of graphene under magnetostatic field bias can lead to a wide variety of nonreciprocal plasmonic components.[173,174,176] For instance, high nonreciprocal isolation was reported with a device consisting of two wire grid polarizers placed around a magnetically-biased graphene sheet.[177] This design, with one wire parallel to the x-axis and the other one tilted by −45° with respect to the x-axis, suppressed the need for multiple graphene layers, and exhibited tunable nonreciprocal operation. The Faraday rotation in magnetically-biased large-area graphene monolayers was also experimentally demonstrated at microwave frequencies.[178] In a relevant experiment, a near optimal nonreciprocal isolator was proposed, again based on magnetostatically biased graphene monolayers but now operating at THz frequencies.[179] Atomically thin nonreciprocal optical isolators were also reported for circularly polarized waves by using magnetized graphene monolayers, opening up new possibilities for additional innovation in the design of tunable ultrathin nonreciprocal optical components, such as isolators and circulators.[180]

Despite their unique properties, the aforementioned graphene-based nonreciprocal structures are based on magneto-optical effects, which require lossy and bulky magnetic materials to create a strong external magnetic field bias and break time-reversal symmetry.[172,181] In order to avoid the problems related to the use of magnetic materials, including large size/weight and incompatibility with the integrated circuit technology, nonlinear materials and spatiotemporal modulation have been suggested as alternative approaches to break reciprocity.[182–187] However, nonlinear effects are inherently weak and usually require high-intensity signals to be excited. In addition, the nonlinear approach to nonreciprocity is more difficult to be realized with a compact passive design at higher frequencies than microwaves and always suffers from the "dynamic" reciprocity problem.[182,188,189]

The spatiotemporal modulation of graphene was recently reported as an alternative approach to break time-reversal symmetry without the detrimental need of magnets.[49] Magnet-free nonreciprocal plasmonic devices and antennas were presented by dynamically modulating the chemical potential of graphene following the formula: $\mu_c(z,t) = \mu_{c0}[1 + M cos(\omega_m t - \beta_m z)]$, where $\mu_{c0}$ is the graphene static chemical potential, $M$ is the modulation depth, $t$ is time, $\omega_m$ is the modulation frequency, $\beta_m = 2\pi/p$ is the modulation wavenumber with $p$ being the modulation spatial periodicity, and $z$ is the direction of electromagnetic wave propagation. The time-modulated graphene response was used in the design of the parallel-plate nonreciprocal waveguide shown in **Figure 13**(a), where each graphene sheet was modulated independently through different applied voltage values by using multiple gating pads. The spatiotemporal modulation of chemical potential, caused by the applied time-varying voltage values, yielded a similar modulation in the effective THz conductivity of graphene, which was given by: $\sigma_{eff}(z,t) = \sigma_0[1 + M cos(\omega_m t - \beta_m z)]$, where $\sigma_0$ is the graphene conductivity without modulation. This approach to nonreciprocity allowed to dramatically modify the radiation pattern of leaky wave antennas based on the time-modulated waveguide design. However, its main drawback was the requirement of multiple gate electrodes underneath graphene to obtain the desirable sinusoidal surface reactance that can cause leaky wave radiation. The large number of electrodes is expected to substantially increase the fabrication complexity of this configuration. The follow-up work improved on this issue and proposed the coupling of a time-modulated graphene capacitor, biased by a single time-modulated voltage, to a low loss dielectric photonic waveguide.[190] The relevant design is shown in Figure 13(b), where a pair of closely spaced time-modulated graphene sheets form the capacitor which is located on top of the dielectric waveguide. This hybrid graphene-dielectric time-modulated waveguide exhibited large nonreciprocal response using realistic bias voltage values. This design was also found to be robust to graphene's inherent loss, thus, providing a prospective platform to develop low-loss compact photonic circulators and Faraday rotators that are compatible to the well-established complementary metal-oxide semiconductor (CMOS) fabrication techniques.

## 5. Conclusions and future perspectives

To conclude, this paper presented a comprehensive review of various new cutting-edge graphene-based photonic technologies. Novel nonlinear, active, topological, and nonreciprocal photonic graphene devices were demonstrated that constitute emerging new technological advances in the research field of THz photonics. First, we discussed that nanostructured graphene or its hybridization with plasmonic and dielectric structures or metamaterials can lead to a large enhancement of electric field that will increase the nonlinear graphene-light interactions. These are the two most prominent approaches to further boost the high intrinsic nonlinearity of graphene and make it more practical. Towards this goal, we have reviewed various recent experimental observations of strong third-order nonlinear phenomena, as well as nonlinear wave mixing, and high harmonic generation generated by various graphene structures. The active properties of graphene were also summarized in the case of optical pumping and the physical mechanisms of the resulted negative photoconductivity were explained. In addition, various emerging photonic applications of photoexcited graphene were discussed, including THz lasers, sensors, and modulators. Several topological photonic designs based on structured graphene were also comprehensively demonstrated. The broadband excitation of topologically protected edge plasmons was presented based on either broken time-reversal symmetry due to magnetically biased graphene or preserved time-reversal symmetry by manipulating graphene's doping level, i.e., its plasmonic properties, in specific locations. Finally, several nonreciprocal graphene photonic designs were discussed by biasing graphene with a magnetostatic field or by exploiting the spatiotemporal modulation of graphene. The latter approach provides a promising new platform to develop magnet-free low-loss compact THz photonic nonreciprocal components.

However, despite the tremendous research advancements in graphene photonics, still significant improvements in the large-scale fabrication, reproducibility, and doping performance of graphene are required in order to move this promising technology to industrial level applications. Moreover, many other newly discovered 2D materials exist except of graphene, like transition metal dichalcogenides (TMDs),[191,192] hexagonal boron nitride (h-BN),[193] black phosphorus,[194,195] and gallium selenide (GaSe),[196] to name a few, that demonstrate dielectric, semiconductor, and metallic (plasmonic) optical properties. The plethora of the recently discovered 2D materials can also be used in the design of new tunable ultrathin photonic devices that can complement the currently presented graphene-based configurations. In addition, a wide variety of Van der Waals heterostructures,[197] such as graphene/h-BN, graphene/TMDs, and TMD/h-BN, have just started to be explored in the literature and fabricated. These multilayer ultrathin structures promise to further improve the performance of various integrated photonic devices. In the near future, it is expected that even more 2D materials will be discovered. The integration of these envisioned new 2D materials to novel photonic platforms will further stimulate the development of integrated THz photonics leading to substantially improved and novel functionalities.

**Acknowledgements**

This work was partially supported by the National Science Foundation Nebraska Materials Research Science and Engineering Center (Grant No. DMR1420645) and the Jane Robertson Layman Fund from the University of Nebraska Foundation.

**Figures**

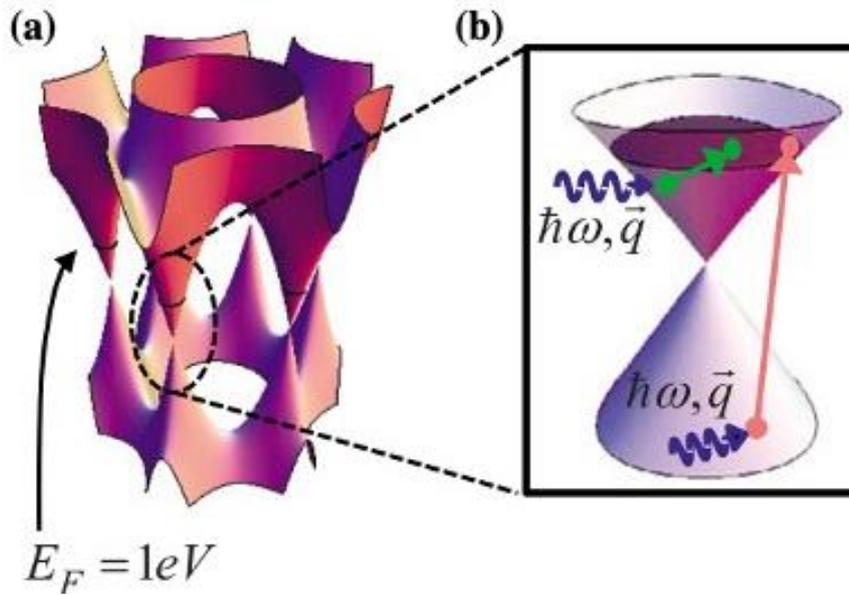

Figure 1. (a) 3D band structure of graphene. The Fermi level of 1 eV is depicted to indicate the vertical scale. (b) Intraband (green arrows) and interband (red arrows) single particle excitations in graphene. Reproduced with permission.[38] Copyright 2009, American Physical Society.)

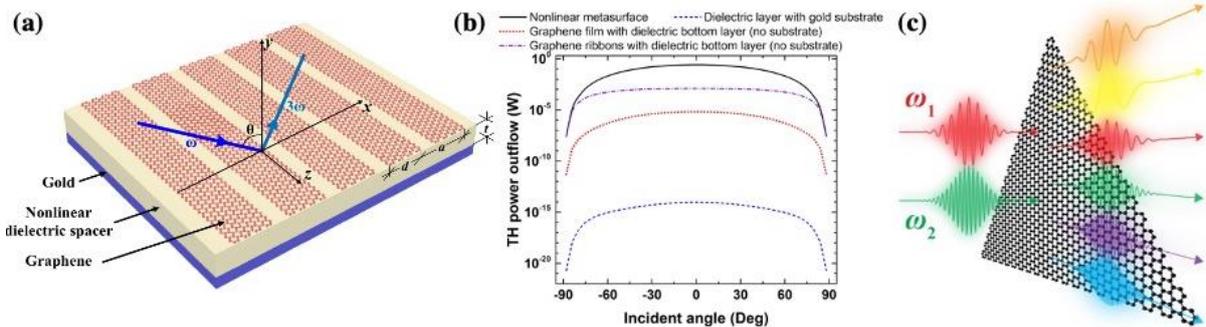

Figure 2. (a) Nonlinear graphene metasurface consisted of an array of graphene ribbons placed on a glass substrate and terminated by a gold reflector. (b) Computed third harmonic power outflow under different scenarios as a function of the incident angle.[91] (c) Plasmon-enhanced wave mixing by doped graphene nanoislands.[95] ((a), (b) Reproduced with permission.[91] Copyright 2017, IOP Publishing Ltd. (c) Reproduced with permission.[95] Copyright 2015, American Chemical Society.)

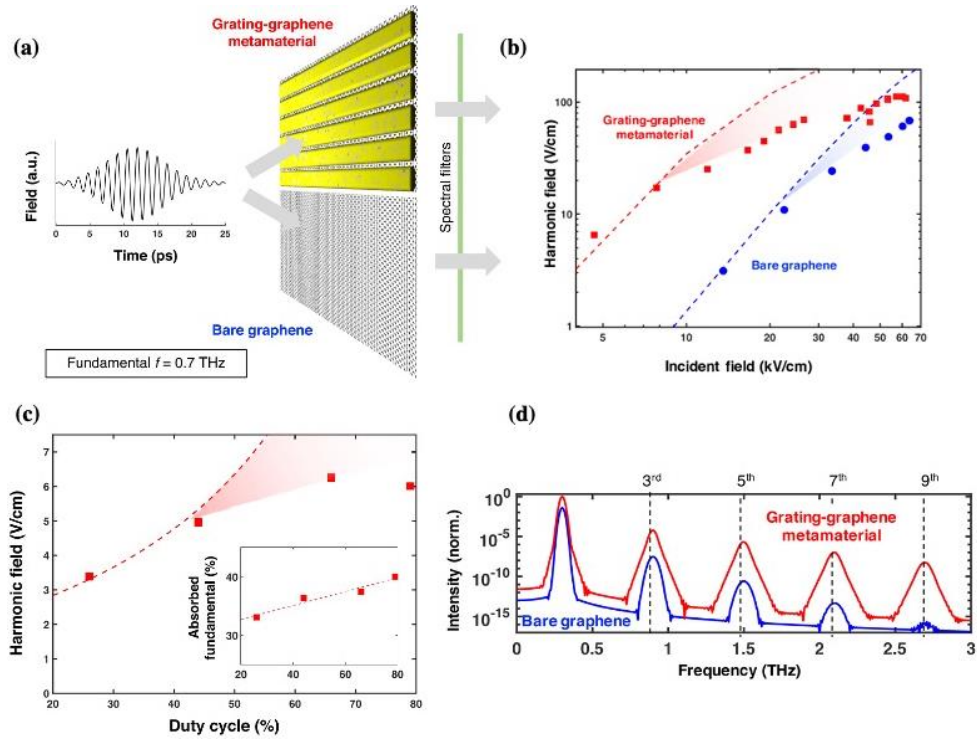

Figure 3. (a) THz measurement setup of a metallic grating-graphene hybrid metamaterial and bare graphene. (b) Third harmonic field intensity as a function of the peak field strength of the incident THz light for the two configurations. (c) Third harmonic field intensity as a function of duty cycle for the grating-graphene hybrid metamaterial. (d) Simulation results of high harmonic intensity generated by the proposed grating-graphene hybrid metamaterial and bare graphene.[97] (Reproduced with permission.[97] Copyright 2021, American Chemical Society.)

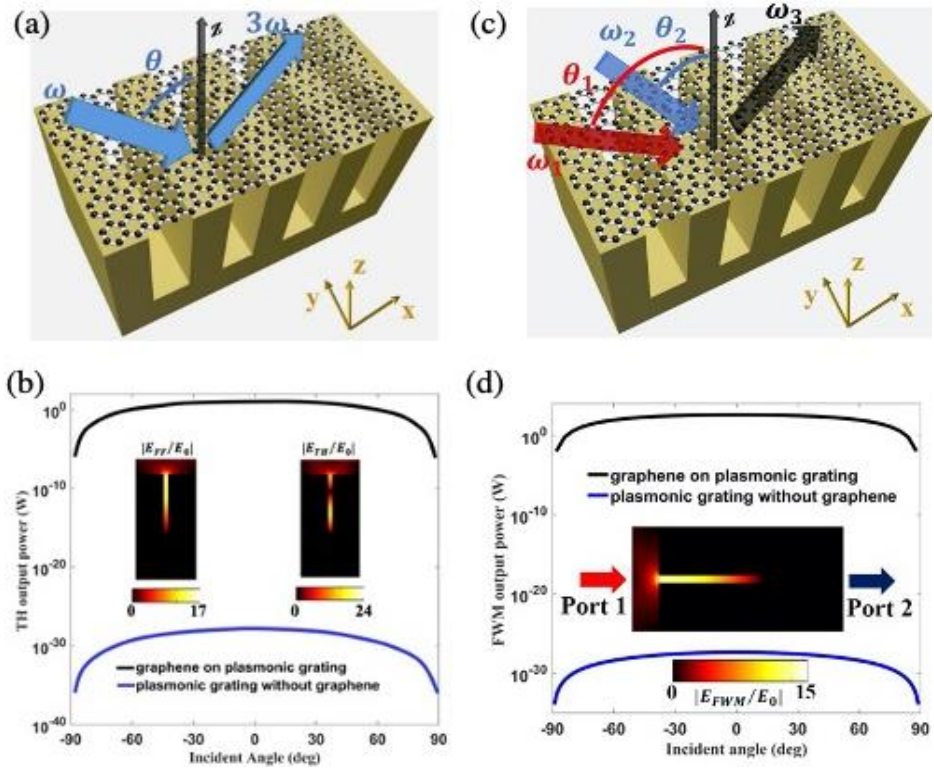

Figure 4. (a) THG process based on a hybrid graphene-plasmonic grating. (b) THG output power by the plasmonic grating with (black line) and without graphene (blue line) on top as a function of the incident angle. (c) FWM process by using the same system shown in (a). (d) FWM output power by the plasmonic grating with (black line) and without graphene (blue line) on top as a function of the incident angle.[98] ((a)-(d) Reproduced with permission.[98] Copyright 2019, American Physical Society.)

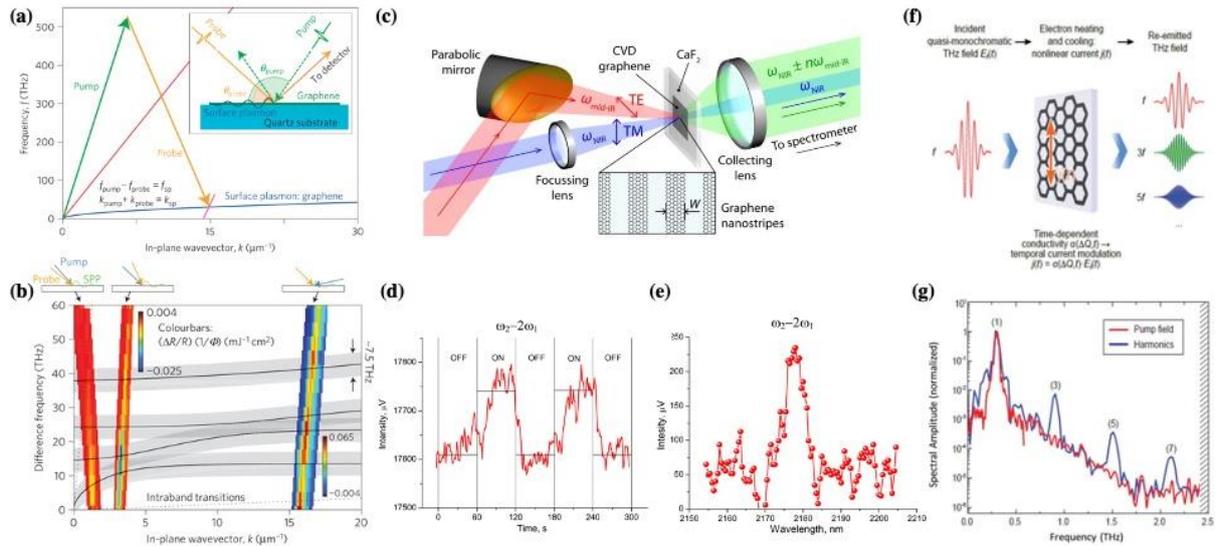

Figure 5. (a) Nonlinear wave mixing to optically generate graphene surface plasmons illustrated by the dispersion diagram of an extended graphene monolayer. (b) Measured differential reflection as a function of the temporal overlap between the pump and probe pulses for the mixing process depicted in (a).[101] (c) Experimental set-up to study the nonlinear wave mixing process generated by graphene nanoribbons. (d)-(e) Measured signal at the mixing frequency $\omega_2 - 2\omega_1$ in (d) time and (e) frequency domain.[93] (f) High harmonic generation from graphene on a substrate driven by a high power THz pulse. (g) Broadband measured transmitted THz spectrum through the graphene sample illustrated in (f) (blue line) showing high harmonic generation relative to the spectrum of the driving field with 0.3 THz fundamental frequency (red line).[70] ((a), (b) Reproduced with permission.[101] Copyright 2016, Nature Publishing Group. (c)-(e) Reproduced with permission.[93] Copyright 2018, American Chemical Society. (f), (g) Reproduced with permission.[70] Copyright 2018, Nature Publishing Group.)

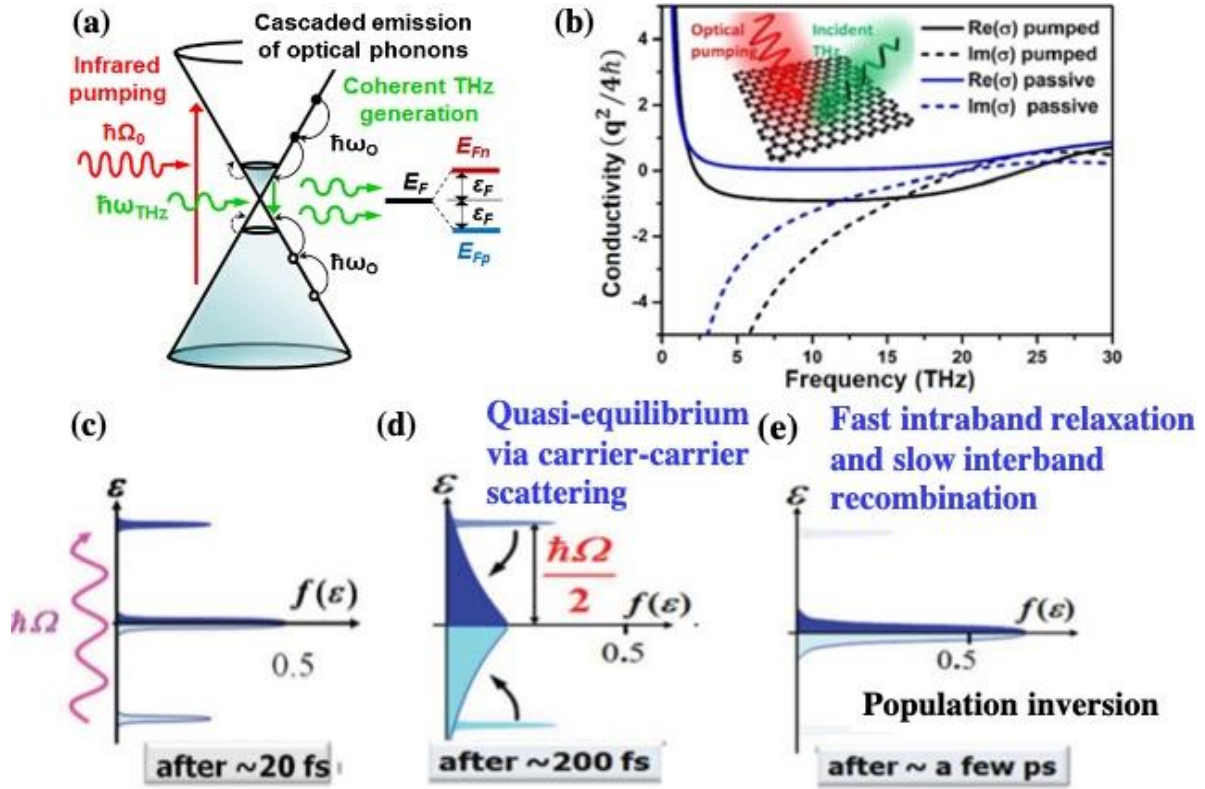

Figure 6. (a) Band structure schematic of THz lasing with optically pumped graphene.[135] (b) The calculated real and imaginary part of conductivity as a function of frequency for pumped (active) graphene and unpumped (passive) graphene.[133] (c) Discrete photogenerated carrier distributions established at the levels $\pm \hbar\Omega/2$ when graphene is optically pumped. (d) Ultrafast quasi-equilibrium reached via carrier-carrier scattering in the femtosecond time scale. (e) Fast intraband relaxation via optical phonons and slow interband electron-hole recombination lead to population inversion in optically pumped graphene after a few picoseconds.[108] ((a) Reproduced with permission. [135] Copyright 2016, American Physical Society. (b) Reproduced with permission. [133] Copyright 2018, Optical Society of America. (c)-(e) Reproduced with permission.[108] Copyright 2012, American Physical Society.)

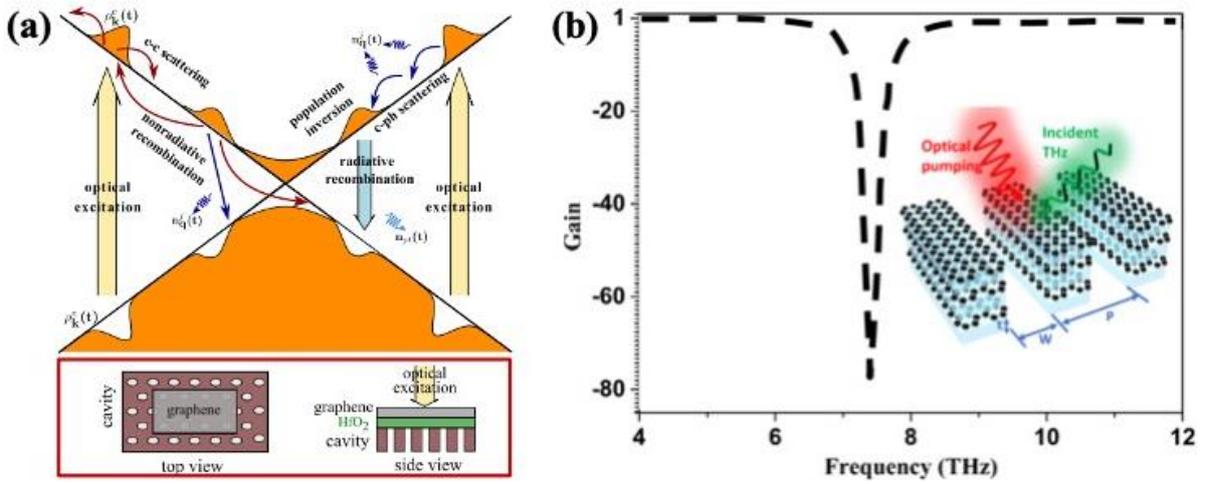

Figure 7. (a) Top: dominant mechanisms in optically pumped graphene. Bottom inset: top and side views of envisioned broadband THz lasers consisting of graphene integrated within a planar photonic crystal cavity.[126] (b) Computed gain response obtained by a patterned hyperbolic metamaterial (shown in the inset) composed of multiple photoexcited graphene sheets stacked between dielectric layers.[133] ((a) Reproduced with permission.[126] Copyright 2015, American Physical Society. (b) Reproduced with permission.[133] Copyright 2018, Optical Society of America.)

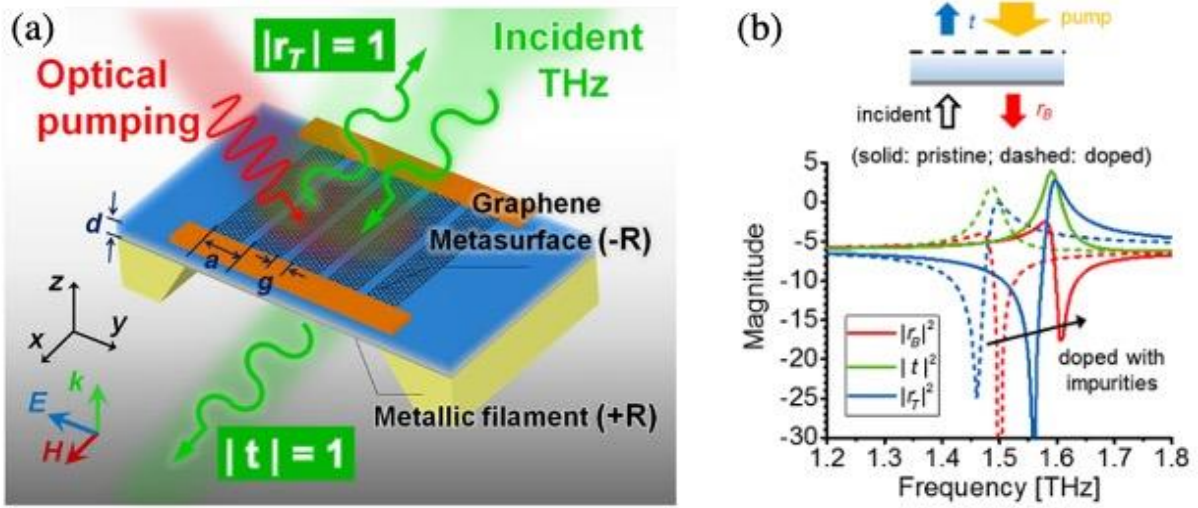

Figure 8. (a) PT-symmetric active THz sensor consisted of an optically pumped graphene metasurface and a metallic filament. (b) Transmission and back/front reflection performance of the graphene-based PT-symmetric sensor before ($\varepsilon_F = 0 meV$, dashed line) and after ($\varepsilon_F = 5 meV$, solid line) being chemically doped. Reproduced with permission.[135] Copyright 2016, American Physical Society.

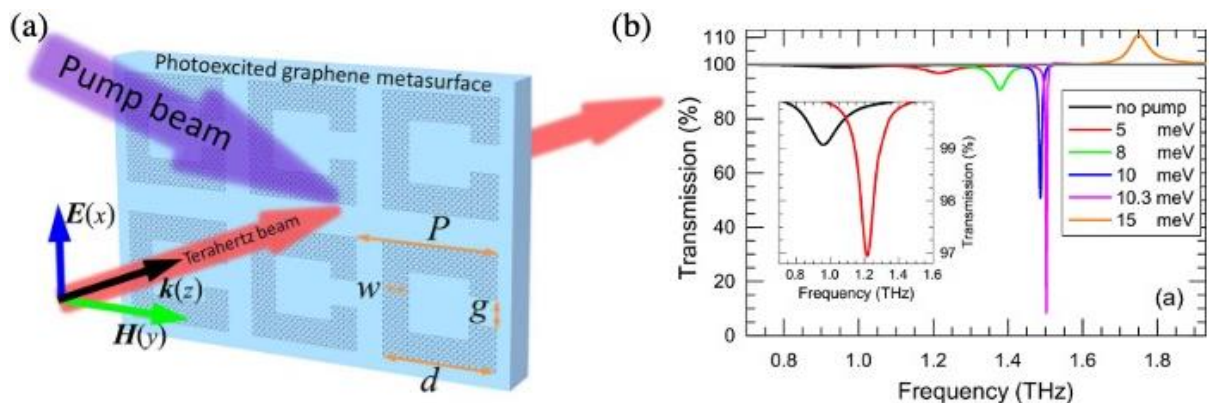

Figure 9. (a) Active graphene metasurface made of a periodic array of photoexcited graphene SRRs over a dielectric substrate. (b) Transmission spectra of the photoexcited graphene metasurface by applying different quasi-Fermi levels. Reproduced with permission.[137] Copyright 2018, American Chemical Society.

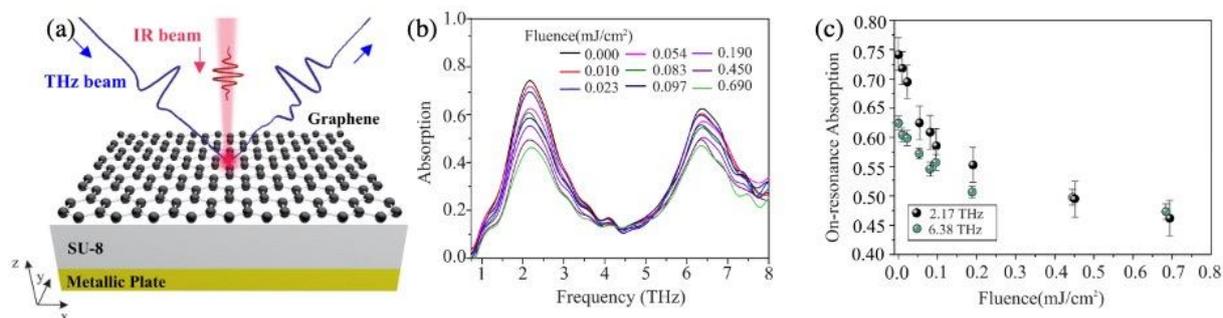

Figure 10. (a) Photoexcited graphene absorber consisting of a graphene monolayer on top of a dielectric (SU-8) film terminated by a metallic plate. The graphene layer is optically excited by an IR laser beam at normal incidence. (b) Absorption modulation for variable fluence values (0-0.690 mJ/cm$^2$) measured over a wide frequency range (0.75-8 THz). (c) Resonant absorption amplitudes as a function of the IR pump fluence. Reproduced with permission.[138] Copyright, 2019, American Chemical Society.

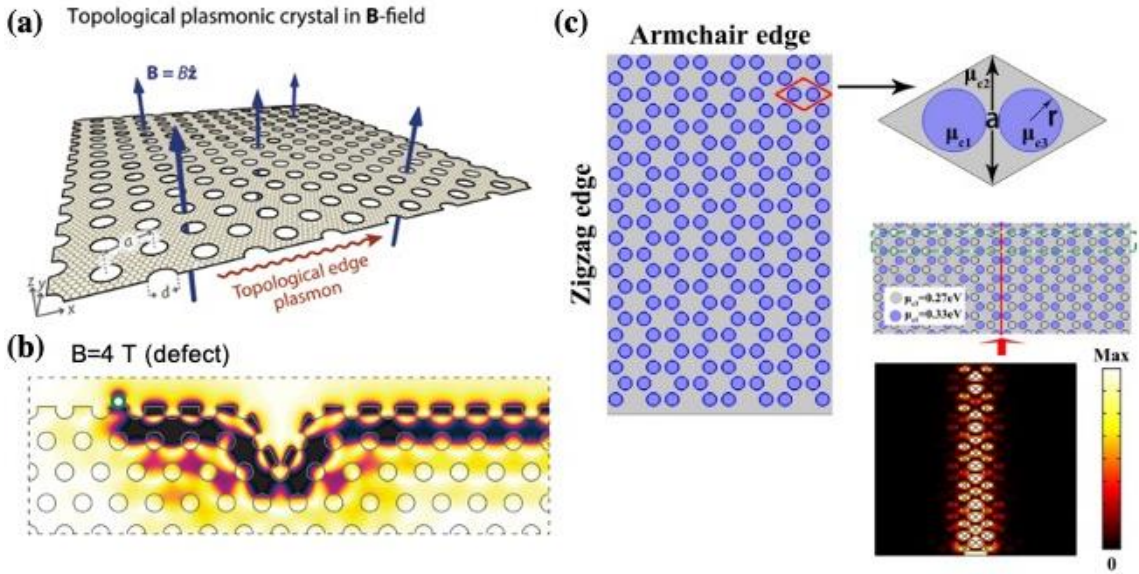

Figure 11. (a) Patterned graphene photonic lattice with broken time-reversal symmetry due to an external magnetic field bias. (b) Topologically protected edge plasmons propagating unperturbed along a structural defect on the design depicted in (a).[164] (c) Graphene plasmonic crystal with honeycomb lattice, where different doping levels are applied to neighboring graphene nanodisks. The electric field distribution of the topologically protected plasmon mode is also shown on the bottom.[165] ((a)-(b) Reproduced with permission.[164] Copyright 2017, American Physical Society. (c) Reproduced with permission.[165] Copyright 2017, American Physical Society.)

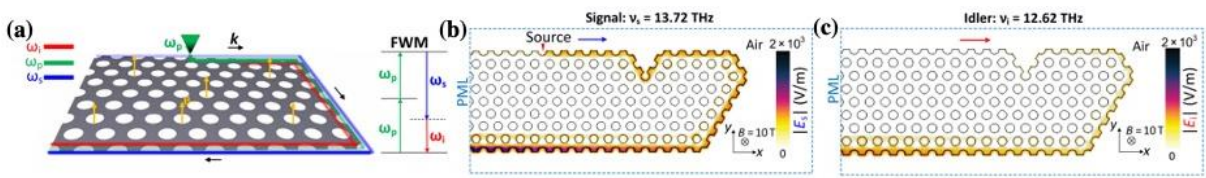

Figure 12. (a) Nonlinear graphene metasurface to achieve topologically protected FWM of edge plasmons. (b)-(c) Field profiles of the edge modes at the signal and idler frequencies, respectively. Reproduced with permission.[171] Copyright 2020, American Association for the Advancement of Science.

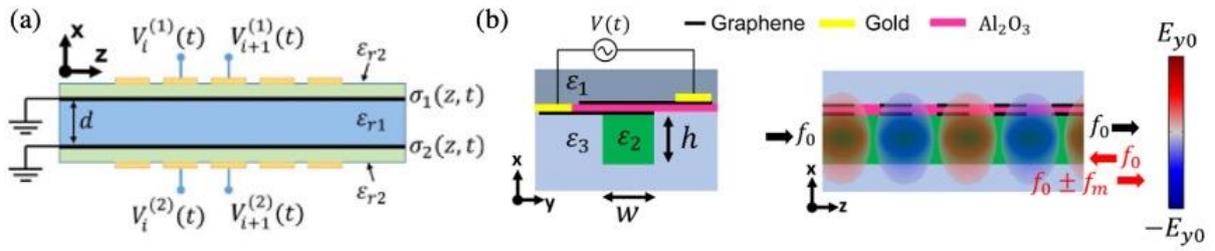

Figure 13. (a) Spatiotemporally modulated graphene to design a tunable parallel-plate waveguide.[49] (b) Dielectric waveguide combined with a time-modulated graphene capacitor (left) and the resulted low-loss photonic isolator design (right).[190] ((a) Reproduced with permission.[49] Copyright 2015, IEEE. (b) Reproduced with permission.[190] Copyright 2018, American Physical Society.)